\documentclass[12pt,preprint]{aastex}
\newcommand{\etal}{\mbox{\rm et al.~}}
\newcommand{\ms}{\mbox{m s$^{-1}~$}}
\newcommand{\ks}{\mbox{km s$^{-1}~$}}

\newcommand{\kse}{\mbox{km s$^{-1}$}}
\newcommand{\mse}{\mbox{m s$^{-1}$}}
\newcommand{\msune}{M$_{\odot}$}

\newcommand{\msun}{M$_{\odot}~$}
\newcommand{\lsun}{L$_{\odot}~$}
\newcommand{\rsun}{R$_{\odot}~$}

\newcommand{\rsune}{R$_{\odot}$}
\newcommand{\mjup}{M$_{\rm JUP}~$}

\newcommand{\msini}{$M \sin i~$}
\newcommand{\vsini}{$v \sin i~$}

\newcommand{\chisq}{$\sqrt{\chi_{\nu}^2}~$}
\newcommand{\arel}{$a_{\rm rel}~$}
\newcommand{\teff}{$T_{\rm eff}~$}

\newcommand{\fe}{{\rm [Fe/H]}}
\newcommand{\logg}{${\rm \log g}~$}
\newcommand{\rhk}{$\log R^\prime_{HK}~$}
\newcommand{\shk}{$S_{HK}~$}
\newcommand{\prot}{$P_{ROT}~$}
  
\received{}
\accepted{}
  
\slugcomment{to appear in ApJ}
  
\shortauthors{Fischer \etal}
\shorttitle{Five Planets}
\begin{document}
  
\title{Five Intermediate-Period Planets from the N2K Sample \altaffilmark{1,2}}
\author{Debra A. Fischer\altaffilmark{3},
Steven S. Vogt\altaffilmark{4},
Geoffrey W. Marcy\altaffilmark{5},
R. Paul Butler\altaffilmark{6},
Bun'ei Sato\altaffilmark{7},
Gregory W. Henry\altaffilmark{8},
Sarah Robinson\altaffilmark{4},
Gregory Laughlin\altaffilmark{4},
Shigeru Ida\altaffilmark{7},
Eri Toyota\altaffilmark{9}
Masashi Omiya\altaffilmark{10}
Peter Driscoll\altaffilmark{11},
Genya Takeda\altaffilmark{12},
Jason T. Wright\altaffilmark{5},
John A. Johnson\altaffilmark{5}}
  
\email{fischer@stars.sfsu.edu}
  
\altaffiltext{1}{Based on observations obtained at the W. M. Keck Observatory,
which is operated by the University of California and the California Institute of
Technology. Keck time has been granted by NOAO and NASA.}
  
\altaffiltext{2}{Based on observations obtained at the Subaru Telescope, which 
is operated by the National Astronomical Observatory of Japan}

\altaffiltext{3}{Department of Physics \& Astronomy, San Francisco State University,
San Francisco, CA  94132; fischer@stars.sfsu.edu}
  
\altaffiltext{4}{UCO/Lick Observatory, University of California at Santa Cruz,
Santa Cruz, CA 95064}

\altaffiltext{5}{Department of Astronomy, University of California,
Berkeley, CA USA 94720}
  
\altaffiltext{6}{Department of Terrestrial Magnetism, Carnegie Institute of
Washington DC, 5241 Broad Branch Rd. NW, Washington DC, USA 20015-1305}
  
\altaffiltext{7}{Tokyo Institute of Technology,
2-12-1 Okayama, Meguro-ku, Tokyo 152-8550, Japan}
  
\altaffiltext{8}{Center of Excellence in Information Systems, Tennessee State
University, 3500 John A. Merritt Blvd., Box 9501, Nashville, TN  37209}
  
\altaffiltext{9} 
{Department of Earth and Planetary Sciences, Graduate School of Science,
Kobe University, 1-1 Rokkodai, Nada, Kobe 657-8501, Japan}

\altaffiltext{10}
{Department of Physics, Tokai University, 1117 Kitakaname,
Hiratsuka, Kanagawa 259-1292, Japan}

\altaffiltext{11}{Department of Earth and Planetary Sciences, Johns Hopkins University,
Baltimore, MD 21218}

\altaffiltext{12}{Department of Physics and Astronomy, Northwestern 
University, 2145 Sheridan Road, Evanston, IL 60208}

\begin{abstract}
We report the detection of five Jovian mass planets 
orbiting high metallicity stars. Four of these stars were first 
observed as part of the N2K program and exhibited low RMS velocity 
scatter after three consecutive observations.  However, follow-up observations 
over the last three years now reveal the presence of 
longer period planets with orbital periods ranging from 21 days 
to a few years.  HD~11506 is a G0V star with a planet of \msini = 4.74 \mjup 
in a 3.85 year orbit. HD~17156 is a G0V star with a 3.12 \mjup planet 
in a 21.2 day orbit.  The eccentricity of this orbit is 0.67,  
one of the highest known for a planet with a relatively short period. 
The orbital period for this planet places it in a region of parameter 
space where relatively few planets have been detected. 
HD~125612 is a G3V star with a 
planet of \msini = 3.5 \mjup in a 1.4 year orbit. HD~170469 is a G5IV star with a planet
of \msini = 0.67 \mjup in a 3.13 year orbit. HD~231701 is an F8V star 
with planet of 1.08 \mjup in a 142 day orbit. All of these stars have
supersolar metallicity.  Three of the five stars were observed photometrically
but showed no evidence of brightness variability.  A transit search conducted
for HD~17156 was negative but covered only 25\% of the search space and so
is not conclusive.
\end{abstract}
  
\keywords{planetary systems -- stars: individual (HD 11506, HD 17156,
HD 125612, HD 170469, HD 231701)}

\section{Introduction}

Ongoing Doppler surveys of stars closer than 150 pc have 
detected more than 200 exoplanets (Butler \etal 2006, Wright \etal 2007). 
This ensemble of exoplanets exhibit a diverse range of statistical characteristics
(Marcy \etal 2005).  Notably, the mass distribution of exoplanets 
falls exponentially for masses greater than one Jupiter-mass. In addition, there is a 
statistical pile-up of planets in orbits of just a few days, a paucity of 
planets with periods between 10 - 100 days, and a rising number of gas giant 
planets found at separations greater than 1 AU.  Orbital eccentricities 
span a surprising range from 0 - 0.93, although 92\% of planet eccentricities
are less than 0.6 and even for planets with periods longer than five days 
(i.e., not tidally circularized) the median exoplanet eccentricity is 0.26. 

It has also been shown that planet formation is tied to the chemical composition 
of the host star. There is a few percent probability of finding a gas 
giant planet around a solar metallicity star, while 
planet occurrence rises dramatically to $\sim 25$\% for stars with three times 
the heavy metal composition of the Sun (Santos \etal 2005, Fischer \& Valenti 2005).  
Ida \& Lin (2004) have accounted for this metallicity correlation within 
the context of core accretion. 

The statistical characteristics of exoplanets serve as tracers of planet 
formation and migration histories. The planet-metallicity correlation 
indicates initial high metallicity in the protoplanetary disk which in 
turn may be correlated with a higher surface density of solid particles 
in the midplane of the disk that enhances core accretion. Orbital eccentricities and the 
proximity of gas giant plants to their host stars are remnant signatures of 
gravitational interactions that drive orbital migration. The architecture
of multi-planet systems, sometimes locked in resonances, adds to our understanding
of evolution of the protoplanetary disk. 

The N2K program (Fischer \etal 2005) is a survey of metal-rich stars, 
designed to identify short period planets. These planets are geometrically 
endowed with a higher transit probability; transit events provide a rare opportunity 
to derive information about the planet density, internal structure and the 
atmosphere (Burrows \etal 2007, Charbonneau 2006, Sato \etal 2005).  
Because short-period planets can be flagged with just a few observations, 
the N2K program only requires three Doppler measurements to screen each star. 
However, an increased occurrence of planets is correlated with high 
host star metallicity at all detected separations. Therefore, additional 
observations were obtained for the highest metallicity stars to check for longer
period planets. This extended program has detected six new intermediate 
period planets: HD~5319 and HD~75898 (Robinson \etal 2007), and four of the 
five planets presented here: HD~11506, HD~17156, HD~125612, HD~231701.

\section{HD 11506}
  
\subsection{Stellar Characteristics}

HD 11506 is classified as a G0 star with V=7.51. The Hipparcos catalog (ESA 1997) 
lists B-V = 0.607 with a parallax of 18.58 milliarcseconds, corresponding to a distance of 
53.8 pc. The distance and apparent magnitude set the absolute visual magnitude 
as $M_V$=3.85 and the stellar bolometric luminosity 
as 2.29 \lsun, including a bolometric correction of -0.033 
(VandenBerg \& Clem 2003) based on effective temperature, surface gravity and 
metallicity of the star. High resolution spectroscopic analysis 
described in Valenti \& Fischer (2005) yields 
\teff = 6058 $\pm$ 51K, \logg = 4.32 $\pm$ 0.08,
\vsini = 5.0 $\pm$ 0.5 \ks, and \fe = 0.31 $\pm$ 0.03 dex.
HD 11506 is about 0.65 magnitudes above the main sequence.  We expect the 
star to be located about 0.3 magnitudes above the main sequence because of  
high stellar metallicity, so this star appears to be slightly evolved 
by a few tenths of a magnitude, and is likely just 
beginning to transition onto the subgiant branch. 

The stellar radius is calculated to be 1.3 \rsun using
$L = 4 \pi R^2  \sigma T^4$.
We have also run a fine grid of evolutionary tracks (described in Takeda \etal 2007), 
tuned to the uniform spectroscopic analysis of Valenti \& Fischer (2005) and 
based on the Yale Stellar Evolution Code.  This analysis provides posterior 
probability distributions for stellar mass, radius, gravity and ages.  Based 
on these evolutionary tracks, we derive a stellar mass of 1.19 \msune,
a radius of 1.38 \rsune, and an age of 5.4 Gyr.  As a measure of the 
formal uncertainties, the lower and upper 95\% credibility intervals from 
a Bayesian posterior probability distribution are 
provided in parentheses in Table~1 for these values.

The Ca II H \& K lines (Figure 1) show that HD 11506 is chromospherically inactive.
We measure \shk, the core emission in the Ca II H \& K lines relative to the 
continuum, for all of our stars. 
Based on nineteen Keck observations, we measure an average \shk = 0.156 
for HD~11506.  The ratio of flux from \shk to the bolometric stellar flux is 
designated as \rhk and gives the best diagnostic of chromospheric activity. 
The average \rhk$= -4.99$ for this star and we derive an activity-based rotational
period (Noyes \etal 1984), \prot$ = 12.6 d$, and an 
activity-based age of 5.4 Gyr, 
in excellent agreement with the age from evolutionary tracks. 
The activity, spectral type and evolutionary stage of the star allow us to estimate 
an additional source of astrophysical noise, or stellar jitter for the velocities of each
of the stars on our Doppler survey (Wright 2005). 

We monitored the brightness of HD~11506 with the T10 0.8~m automatic 
photometric telescope (APT) at Fairborn Observatory (Henry 1999, 
Eaton, Henry \& Fekel 2003).  The T10 APT measures 
the brightness of program stars relative to nearby constant comparison stars 
with a typical precision of 0.0015--0.0020 mag for a single measurement.  
For HD~11506, we obtained 102 $b$ and $y$ measurements spanning 451 days 
between 2004 October and 2006 January.  The standard deviation of a single 
observation from the mean was 0.0023 mag, our upper limit to possible 
photometric variability in HD~11506.  A periodogram analysis found no 
significant periodicity between 1 and 225 days, so our photometry confirms 
the star's low chromospheric activity.  The stellar parameters are summarized 
in Table~1.

\subsection{Doppler Observations and Keplerian Fit}
  
Doppler observations were made at the Keck telescope using HIRES 
(Vogt \etal 1994) with 
an iodine cell to model the instrumental profile and to provide the wavelength scale 
(Butler \etal 1996). An exposure meter maintains a constant signal-to-noise ratio 
of about 200 in our spectra, yielding a mean radial velocity precision of 2.75 \ms for 
HD~11506. We obtained a total of 26 Doppler measurements. The observation 
dates, radial velocities and measurement uncertainties for the 
radial velocities are listed in Table 2 
and plotted in Figure 2. 

In addition to velocity errors arising from our measurement uncertainties (including 
photon shot noise), the star itself can have pulsations, cool spots or granular convective 
flows that contribute non-dynamical velocity noise. These astrophysical sources of noise 
are termed jitter and we empirically estimate stellar jitter based on the spectral type and 
chromospheric activity of the star, following Wright (2005). 
For purposes of fitting a Keplerian model, the stellar jitter is added 
in quadrature to the formal instrumental errors, however the estimated jitter is never 
included in the measurement uncertainties for the tabulated radial velocity sets. 

The periodogram of the radial velocities shows a strong, broad peak in the power 
spectrum at about 1270 days with an associated false alarm probability (FAP) $< 0.0001$. The FAP 
associated with the periodogram tests whether scrambled velocities yield power that
exceeds the observed, unscrambled velocities.  A high FAP suggests that 
the signal is not significant or could have been caused by a window function
in the data. Using a Monte Carlo simulation, one thousand data sets 
of noise were generated by randomly drawing (with replacement) sets of actual stellar 
velocities. The fraction of trials with maximum periodogram power 
that exceeds the observed value from the initial unscrambled 
data set defines the FAP (Cumming 2004). 

For each of the radial velocity data sets in this paper, a 
Levenberg-Marquardt fitting algorithm was used to model the 
radial velocities with a theoretical Keplerian orbital curve.   There are six orbital 
parameters derived in the fit: orbital period (P), time of periastron passage ($T_P$), 
eccentricity (e), the orientation of the orbit (or line of apsides) ($\omega$), the 
semi-velocity amplitude (K) and the residual center of mass radial velocity 
(after subtracting a median radial velocity). 

Uncertainties in the orbital parameters are determined with a bootstrap Monte Carlo 
analysis.  First, a best-fit Keplerian model is obtained. Then, for each of 100 trials, the 
theoretical best fit is subtracted from the observed radial velocities.  The residual 
velocities are then scrambled (with replacement) and added back to the theoretical best fit 
velocities and a new trial Keplerian fit 
is then obtained. The standard deviation of each orbital 
parameter for the 100 Monte Carlo trials was adopted as the parameter uncertainty.
  
The best fit Keplerian model gives an orbital period of 1405 $\pm$ 45 d,
semi-velocity amplitude of 80 $\pm$ 3 \mse and orbital eccentricity of 0.3 $\pm$ 0.1. 
The RMS to this fit is 10.8 \mse. Based on the chromospheric activity of this star,
we estimated a jitter of 2 \ms (Wright 2005).  When this jitter is added in quadrature
with the error bars listed in Table 2, \chisq = 3.2. While the large amplitude 
Doppler variation is clear, the \chisq fit is worse than usual, suggesting 
that our velocity errors may be underestimated or that  
additional low amplitude dynamical velocities are present. A periodogram of 
the residual velocities to the single Keplerian fit show several peaks with 
similar power.  For example, we can fit a second planet with a period of 170 days with a 
significant reduction in the residual velocity RMS and an improvement in \chisq, 
however this is not yet a unique double planet fit; additional data are required to better
evaluate the possible second signal. 

Using the stellar mass of 1.19 \msun derived from 
evolutionary tracks, we find \msini = 4.74 \mjup and a semi-major axis of 2.48 AU. 
At the distance of this star, this physical separation corresponds to an 
angular separation of $\alpha = 0.'' 04$.  
The Keplerian orbital solution is listed in Table 3 and the best-fit 
Keplerian model is plotted in Figure 2.  

\section{HD 17156}
  
\subsection{Stellar Characteristics}
  
HD 17156 is a listed as a G5 star in the SIMBAD database and the 
Hipparcos catalog.  However, this spectral type seems at odds with other data 
for the star.  The visual magnitude is $V = 8.17$, 
$B-V = 0.59$, and the Hipparcos parallax (ESA 1997) is 12.78 milliarcseconds, 
corresponding to a distance of 78.24 pc. The bolometric correction -0.039 
(VandenBerg \& Clem 2003) and absolute visual magnitude, $M_V$=3.70,  
imply a bolometric stellar luminosity of 2.6 \lsun.  Spectroscopic analysis 
yields \teff = 6079 $\pm$ 56K, \logg = 4.29 $\pm$ 0.06, \vsini = 2.6 $\pm$ 0.5 \ks, and 
\fe = 0.24 $\pm$ 0.03. The $B - V$ color and the effective temperature 
are independent measurements that are consistent with each other.  
Together with the absolute 
magnitude and position on the H-R diagram, the spectral type for this star is 
more likely to be G0 and the star is just beginning to evolve off the main sequence.

The stellar mass, from evolutionary models described by 
Takeda \etal (2007), is 1.2 \msune, and the age is 5.7 Gyr.  
The stellar radius from evolutionary models is 1.47 \rsune, and agrees with 
the value we derive using the observed luminosity and the Stefan-Boltzmann relation. 

The absence of Ca II H \& K emission (Figure 1) demonstrates low chromospheric activity. 
Taking the average of 25 observations, we measure \shk = 0.15 and \rhk = -5.04 and derive 
a rotational period, \prot = 12.8 d, with an estimated stellar age of $6.4 \pm 2$Gyr, 
which compares favorably with the age derived above from stellar evolution tracks. 

We obtained 241 photometric measurements with the T12 APT spanning 179 days 
between 2006 September and 2007 March.  The standard deviation of the 
observations from their mean was 0.0024 mag, the upper limit to photometric 
variability in the star. Periodogram analysis revealed no significant 
periodicity between 1 and 100 days.  In particular, a least-squares sine fit 
of the observations on the 21.22-day radial velocity period resulted in a 
photometric amplitude of only $0.00039 \pm 0.00023$ mag, providing further 
evidence that the radial velocity variations in HD~17156 are not due to 
chromospheric activity.  The stellar characteristics, including our assessment 
of photometric variability, are summarized in Table 1.

\subsection{Doppler Observations and Keplerian Fit}

We initially obtained eight Doppler observations of HD 17156 using the High Dispersion 
Spectrometer (Noguchi \etal 2002) at the Subaru Telescope in 2004 and 2005.  
For the first observing runs, the iodine absorption cell was located behind 
the entrance slit of the spectrometer (Kambe \etal 2002, Sato \etal 2002, Sato \etal 2005).
The box holding the I2 cell included a window with a lens to maintain constant focal 
length inside the spectrometer.
This eliminated the need to adjust the collimator position when 
moving the I2 cell in and out of the light path (i.e., when taking program 
and template observations).
However, the lens introduced a different wavelength dispersion 
for program observations relative to the template observation. 
Modeling of those early data is still ongoing, however, 
standard stars, known to have constant radial velocities show RMS 
scatter greater than 15 \mse, with larger run-to-run velocity offsets 
for Doppler observations obtained with that setup. 

The Subaru N2K program was awarded ten nights of ``intensive'' time in summer 2006 and in 
December 2006. Before the intensive time allocation, the iodine cell was moved in front of 
the slit, eliminating the change in wavelength dispersion between 
template and program observations.
With this new setup, the RMS scatter decreased, 
ranging from 4 - 12 \ms in a set of four RV standard stars.

HD 17156 had exhibited large radial velocity variations in 2004 - 2005 at Subaru. 
Follow-up observations at Keck confirmed velocity variations, so the star was observed on nine 
consecutive nights at Subaru from 8 December to 16 December, 2006. Setup StdI2b was 
used to cover the wavelength region of 3500--6100 \AA \ with a mosaic of 
two CCDs. The slit width of 0$^{\prime\prime}$.6 was used to 
give a reciprocal resolution ($\lambda/\Delta\lambda$) of 60000. We obtained a typical 
signal-to-noise ratio of $S/N\sim150$ pixel$^{-1}$ at 5500 \AA \ with 
exposure times of about 120 seconds. Because of the larger systematic errors for
observations taken before summer 2006 (with the iodine cell behind the slit), only 
the nine radial velocities from December 2006 are listed in Table 4.  To account
for the intrinsic RMS velocity scatter in standard stars, 5 \ms was added in 
quadrature to the nine Subaru observations in Table 4. 

After HD 17156 was flagged as an N2K candidate at Subaru, it was added 
to the N2K planet search program at Keck. We obtained 24 radial velocity 
measurements at the Keck Observatory with an average 
internal velocity precision of 1.6 \mse. 
Observation dates, radial velocities and uncertainties for 33 observations are listed in Table 4.
The last column designates the source of the observations as ``K'' (Keck Observatory) 
or ``S'' (Subaru Observatory). The periodogram of the radial velocity data shows 
a strong narrow peak at 21.1 days with a FAP less than 0.0001 (for 10000 Monte Carlo trials).

When combining the Subaru and Keck velocities, we first determined a velocity difference
of about 130 \ms between two observations taken at Subaru and Keck on the same night (JD 2454083.9).
With that initial guess, we included a velocity offset as a free parameter and found 
that an offset of 116.0 \ms produced a minimum \chisq. 
That offset was added to the Subaru velocities listed in Table 4. 
The best fit Keplerian model for the combined Subaru and Keck data sets 
yields an orbital period of 21.2 $\pm$ 0.3 d, 
semi-velocity amplitude, $K = 275 \pm$ 15 \mse, 
and orbital eccentricity, $e = 0.67 \pm$ 0.08.  
The RMS to the fit is 3.97 \mse.  Adding jitter of 3 \ms (expected for this star) 
in quadrature with the actual 
single-measurement errors gives \chisq = 1.04 for this Keplerian fit. 

Adopting a stellar mass of 1.2 \msune, we derive \msini = 3.12 \mjup and a semi-major axis
of 0.15 AU.  The Keplerian orbital solution is summarized in Table 3. The phase-folded plot of 
the Doppler measurements for Keck and Subaru observations are shown in the left plot of Figure 3 and 
include 3 \ms jitter.  Keck observations are represented by diamonds and the Subaru 
observations are shown as filled circles.  

Because the high eccentricity is unusual, we examined the Keplerian fit for the 
Keck data alone, shown in the right plot Figure 3. The Keck data have 
poor phase coverage near periastron, and yield a Keplerian 
fit with lower amplitude and lower eccentricity. The Subaru observations map
periastron passage and help to model the eccentricity 
of the orbit.

\subsection{Transit Search}
    
The 21.22 day period of the companion to HD~17156 is by far the shortest 
planetary orbital period in this paper.  The orbital semi-major axis of 
0.15 AU and the stellar radius of 1.47 \rsun lead to an {\it a priori} 
transit probability of 7\% (Seagroves et al. 2003).  Therefore, 
we used our 241 brightness measurements to conduct a preliminary transit 
search.  The orbital parameters in Table~3 constrain the predicted times 
of transit to about $\pm~0.3$ days, which is slightly greater than the 
0.25-day duration of a central transit.  We performed our transit search, 
using a technique similar to the one described by Laughlin (2000), over 
all orbital phases for periods between 20 and 23 days.  The search was 
negative but was able to cover effectively only 25\% of the period-phase 
search space corresponding to the uncertainties in the orbital parameters.  
Thus, our photometric data do not preclude the possibility of transits in 
HD~17156.

\section{HD 125612}
  
\subsection{Stellar Characteristics}
  
HD 125612 is a G3V main sequence star with V=8.31, B-V = 0.628, and Hipparcos 
parallax (ESA 1997) of 18.93 corresponding to a distance of 52.82 pc 
and absolute visual magnitude, $M_V$=4.69.  Spectroscopic analysis yields 
\teff = 5897 $\pm$ 40K, \logg = 4.45 $\pm$ 0.05,
\vsini = 2.1 $\pm$ 0.5 \ks, and \fe = 0.24 $\pm$ 0.03 dex.
The bolometric correction is -0.061, giving a stellar luminosity of 1.08 \lsun.  
The luminosity and \teff 
imply a stellar radius of 1.0 \rsune. Within uncertainties, this agrees well 
with the value of 1.05 \rsun determined from stellar evolutionary tracks. We also 
derive a stellar mass of 1.1 \msun from stellar evolution models and an age of 2.1 Gyr. 
 
Figure 1 shows the Ca H line for HD~125612; the lack of emission indicates low 
chromospheric activity for this star.  Taking the mean of 18 observations,
we measure \shk = 0.178 and \rhk = -4.85, and derive \prot = 10.5 d, and 
a stellar age of $3.3 \pm 2$ Gyr (which compares well with the age of 2.1 Gyr from
stellar evolution tracks).
Stellar parameters are summarized in Table~1.

\subsection{Doppler Observations and Keplerian Fit}
  
We obtained 19 Keck velocity measurements for HD 125612 with a typical uncertainty of 
2.2 \mse.  Observation dates, radial velocities 
and instrumental uncertainties in the radial velocities are listed in Table 5. 
A periodogram of the velocities shows a strong broad peak at about 500 days.

The best fit Keplerian model is plotted in Figure 4 and yields a period of 510 $\pm$ 14 d,
with semi-velocity amplitude 90.7 $\pm$ 8 \mse, orbital eccentricity, 0.38 $\pm$ 0.05, 
and a linear trend of 0.037 meters per day.  
Adopting a stellar mass of 1.1 \msune, we derive \msini = 3.5 \mjup and semi-major axis
of 1.2 AU (angular separation, $\alpha = 0''.023$).  The Keplerian orbital solution is 
listed in Table 3 and the RV data are plotted with the best-fit Keplerian 
model (solid line) in Figure 4.

The RMS to the Keplerian fit shown in Figure 4 is 10.7 \mse.  The 
velocity jitter for this star is expected to be about 2 
\mse.  Therefore, the residual RMS is several times the typical error bar, consistent with the 
poor \chisq statistic of 3.56.  A periodogram of the 
residuals to a 510-day planet fit shows 
power near 3.5 d. However, there are several other peaks of 
nearly comparable height, showing that other orbital solutions 
may give similar improvements. 
Thus, while we could fit the residuals with a second Keplerian, the FAP of the peak does 
not yet meet our standards of statistical significance, 
and more data are required for follow up.

\section{HD 170469}
  
\subsection{Stellar Characteristics}
  
HD 170469 is a G5 subgiant star with visual magnitude V=8.21, B-V = 0.677, and 
Hipparcos parallax (ESA 1997) of 15.39 milliarcseconds, corresponding to a distance of 64.97 pc.
The absolute visual magnitude of the star is $M_V$=4.14.
The bolometric correction is -0.072 providing a 
bolometric stellar luminosity of 1.6 \lsun and (with \teff) stellar radius of 1.2 \rsun 
calculated from the luminosity.
Evolutionary tracks provide a stellar mass estimate of 1.14 \msun and stellar radius 
of 1.22 \rsun and age of 6.7 Gyr.  Our spectroscopic analysis
gives \teff = 5810 $\pm$ 44K, \logg = 4.32 $\pm$ 0.06,
\vsini = 1.7 $\pm$ 0.5 \ks, and \fe = 0.30 $\pm$ 0.03 dex.

The Ca H \& K lines (Figure 1) indicate low chromospheric activity.
Taking the mean of thirteen observations, we measure 
\shk = 0.145 and \rhk = -5.06 and derive a rotational period,
\prot = 13.0 d and an activity-calibrated age (Noyes \etal 1984) of $7 \pm 2$ Gyr.

We obtained 215 brightness measurements with the T10 APT spanning 630 days 
between 2005 March and 2006 November.  The standard deviation of the observations 
was 0.0018 mag, the upper limit to photometric variability in HD~170469.  A 
periodogram analysis found no significant periodicity between 1 and 315 days, 
confirming the star's low chromospheric activity. The stellar characteristics are 
summarized in Table 6.

\subsection{Doppler Observations and Keplerian Fit}
  
We obtained 35 Keck velocities for HD 170469 with a mean velocity precision of 
1.6 \mse.  Observation dates, radial velocities 
and instrumental uncertainties in the radial velocities are listed in Table 7. 
A periodogram of the velocities yields very strong power at about 1100 days with a 
FAP less than 0.0001. 
 
The best fit Keplerian model gives an orbital period of 1145 $\pm$ 18 d,
semi-velocity amplitude of 12.0 $\pm$ 1.9 \ms and orbital eccentricity, 0.11 $\pm$ 0.08. 
The RMS to the fit is 4.18 \ms with \chisq = 1.59, including 
the estimated astrophysical jitter of 2.0 \mse. 
Adopting a stellar mass of 1.14 \msune, we derive \msini = 0.67 \mjup 
and a semi-major axis of 2 AU ($\alpha = 0.''03$).  The 
Keplerian orbital parameters are listed in Table 8 and the RV data 
are plotted with the best-fit Keplerian model (solid line) in Figure 5.

\section{HD 231701}
  
\subsection{Stellar Characteristics}
  
HD 231701 is an F8V star with $V = 8.97$, $B-V = 0.539$, and Hipparcos parallax 
(ESA 1997) of 9.22 milliarcseconds, corresponding to a distance of 108.4 pc.  
The absolute visual magnitude is $M_V$=3.79, so this star is beginning to evolve 
onto the subgiant branch.
Spectroscopic analysis yields \teff = 6208 $\pm$ 44K, \logg = 4.33 $\pm$ 0.06,
\vsini = 4.0 $\pm$ 0.5 \ks, and \fe = 0.07 $\pm$ 0.03 dex.
The bolometric correction is -0.037 and bolometric 
luminosity is 2.4 \lsun.  The luminosity and effective temperature yield 
a stellar radius of 1.36 \rsun.  Modeling the stellar evolutionary 
tracks,  we derive a stellar mass of 1.14 \msune, radius of 1.35 \rsune, and 
age of 4.9 Gyr. 
 
The Ca H\&K lines (Figure 1) show that the star has low chromospheric activity.
We measure \shk = 0.159 and \rhk = -5.0 and derive a rotational period,
\prot = 12.2 d and a stellar age of $5.6 \pm 2$ Gyr.
Stellar parameters are listed in Table 6.

\subsection{Doppler Observations and Keplerian Fit}
  
We obtained 17 Keck observations of HD 231701 with mean internal 
errors of 3.2 \mse.  Observation dates, radial velocities 
and measurement uncertainties in the radial velocities are listed in Table 9.
The periodogram of this data set has a FAP of 0.006 for a period near
140 days. 
 
The best fit Keplerian model has an orbital period of 141.6 $\pm$ 2.8 d,
with semi-velocity amplitude 39 $\pm$ 3.5 \ms and orbital eccentricity, 0.1 $\pm$ 0.06.  
The RMS to this fit is 5.9 \mse.  The expected astrophysical jitter for this 
star is 2.2 \mse.  Adding this jitter in quadrature with the error bars listed
in Table 9 yields \chisq = 1.46 for this Keplerian fit. 
Adopting the stellar mass of 1.14 \msune, we 
derive \msini = 1.08 \mjup and semi-major axis of 
0.53 AU.  The Keplerian orbital solution is listed in Table 8 and the phased RV data 
is plotted with the best-fit Keplerian model (solid line) in Figure 6.

\subsection{Discussion}

Here, we present the detection of 5 new exoplanets detected 
with Doppler observations.  
For each of the Keplerian models, we also carried out a 
Markov Chain Monte Carlo (MCMC) analysis to better estimate 
the orbital parameters and their uncertainties following the algorithm 
described by Ford (2003). Unlike the Levenberg-Marquardt algorithm that we
generally use to determine a best fit Keplerian orbit, the MCMC analysis 
provides the full posterior probability density distribution for each 
parameter.  This approach is particularly useful for data sets where 
the Levenberg-Marquardt algorithm can minimize \chisq with a model that 
fits a sparse data set. The MCMC algorithm explores a wider range of 
parameter space because it is not driven solely by \chisq minimization.  
However, MCMC does not explore an exhaustive range of parameter space. For 
example, solutions with very different orbital periods might be missed.
For each of the models presented here, we began with the input parameters
found with Levenberg Marquardt fitting and confirmed that the orbital 
elements were recovered with strongly peaked probability distributions 
using MCMC. 

HD 170469 is a star on the regular planet search at Keck that has a planet of 
\msini = 0.66 \mjup in a $\sim$3 yr orbit with eccentricity 0.23. The host star is 
metal-rich with [Fe/H] = 0.3.  
The remaining four exoplanets were initially part of the N2K program
at Keck. The N2K program targets metal-rich stars for rapid identification of 
short-period planets. The first three radial velocity measurements for 
the stars presented here had RMS scatter
less than 5 \mse (except HD 17156, with initial RMS scatter of 34 \ms), so these 
stars were not candidates for short period planets.  However, a follow-up program to obtain 
Doppler observations on N2K-vetted high metallicity stars with low chromospheric 
activity and low RMS velocity scatter has detected the presence of these longer period planets.

HD 11506 b is a fairly massive planet, with \msini = 4.74 \mjup and a semi-major axis
of 2.5 AU.  This could well constitute the outer edge of a habitable zone location 
for putative rocky moons orbiting the planet, depending on atmospheric properties 
of any moons. The host star has a luminosity 
that is 2.3 times that of the Sun. The eccentricity of this system is 0.3, 
so the temperature at the top of the planet 
atmosphere would change by about 50K between apastron and periastron.

HD 17156 b has a mass of \msini = 3.12 \mjup and an orbital period of 21.2 days, placing it 
in the so-called period valley between 10 and 100 days (Udry \etal 2003), 
where a relatively small fraction of exoplanets have been detected. 
We derive a substantial orbital eccentricity of 
0.67 for HD~17156~b.  At this proximity to the subgiant host star, the planet moves between 0.05 and 0.25 AU,
experiencing temperature changes of a few hundred degrees between periastron and apastron.  
It is possible that these thermal changes could be observed with sensitive IR flux 
measurements from space, even though the planet is not known to transit its host star. 

The distribution of orbital eccentricities for known exoplanets is shown in 
Figure 7.  An upper envelope in the distribution of eccentricities rises steeply 
from periods of a few days to reach the maximum observed eccentricities (for HD 80606 
and HD 20782) at periods of 100 - 1000 days.  Although an orbital eccentricity of 0.67 
seems remarkable for HD 17156 b, given its orbital period of just 21.2 days,  
the eccentricity still falls along the upper edge of the observed eccentricity distribution. 

HD 125612 b has \msini = 3.5 \mjup with a semi-major axis of 1.2 AU.  This 
planet has an eccentricity of 0.38. The planet is carried from 0.47 AU 
at periastron, where the temperature at the top of the atmosphere is 
about 300 K, to about 2.1 AU where the temperature falls below the freezing 
point of water to about 200 K.  A single planet model does not appear to 
adequately describe the velocities of HD 125612 because the RMS to that fit is 
10.7 \mse, yet the star is chromospherically quiet and slowly 
rotating, with \vsini = 2 \kse. This star may well have an additional 
planet orbiting in a relatively short period.

Velocity variations in HD 231701 have been modeled as a planet with 
\msini = 1.08 \mjup with a semi-major axis of 0.53 AU and orbital eccentricity 
of about 0.1.  The MCMC probability distributions for HD 231701 are consistent 
with this Keplerian model, but allow for eccentricity 
solutions that extend to zero. This
analysis alerts us that more RV measurements should be taken 
to better constrain the orbital eccentricity of this system.

We have now obtained three or more Doppler observations for 423 stars at 
Keck Observatory as part of the N2K program.  Spectral synthesis modeling 
has been carried out for all of these stars, and we plot the percentage of stars
with detected planets in each 0.1 dex metallicity bin in Figure 8.  Superimposed on 
this plot is the planet detectability curve from Fischer \& Valenti (2005). 
A planet probability can be assigned based on the stellar metallicity. Integrating
planet probabilities we expect $27 \pm 5$ exoplanets with masses greater than 
1 \mjup and orbital periods shorter than 4 years.  Fourteen, or about half 
of the expected planets in the sample have now been detected.

\acknowledgements
We gratefully acknowledge the dedication and support of the 
Keck Observatory staff, in particular Grant Hill for support with HIRES.
We thank Akito Tajitsu and Tae-Soo Pyo for their expertise and support of the Subaru HDS 
observations. DAF acknowledges support from NASA grant NNG05G164G and from 
Research Corporation.  SSV acknowledges support from NSF AST-0307493. BS is supported
by Grants-in-Aid for Scientific Research (No. 17740106) from the Japan Society 
for the Promotion of Science (JSPS). We thank the Michelson Science Center for 
travel support through the KPDA program.
We thank the NASA and UC Telescope assignment committees for generous 
allocations of telescope time.  
The authors extend thanks to those of
Hawaiian ancestry on whose sacred mountain of Mauna Kea we are
privileged to be guests.  Without their kind hospitality, the
Keck observations presented here would not have been possible.
This research has made use of the SIMBAD database, operated at CDS, 
Strasbourg, France, and of NASA's Astrophysics Data System Bibliographic 
Services and is made possible by the generous support of 
Sun Microsystems, NASA, and the NSF.

\begin{figure}
\plottwo{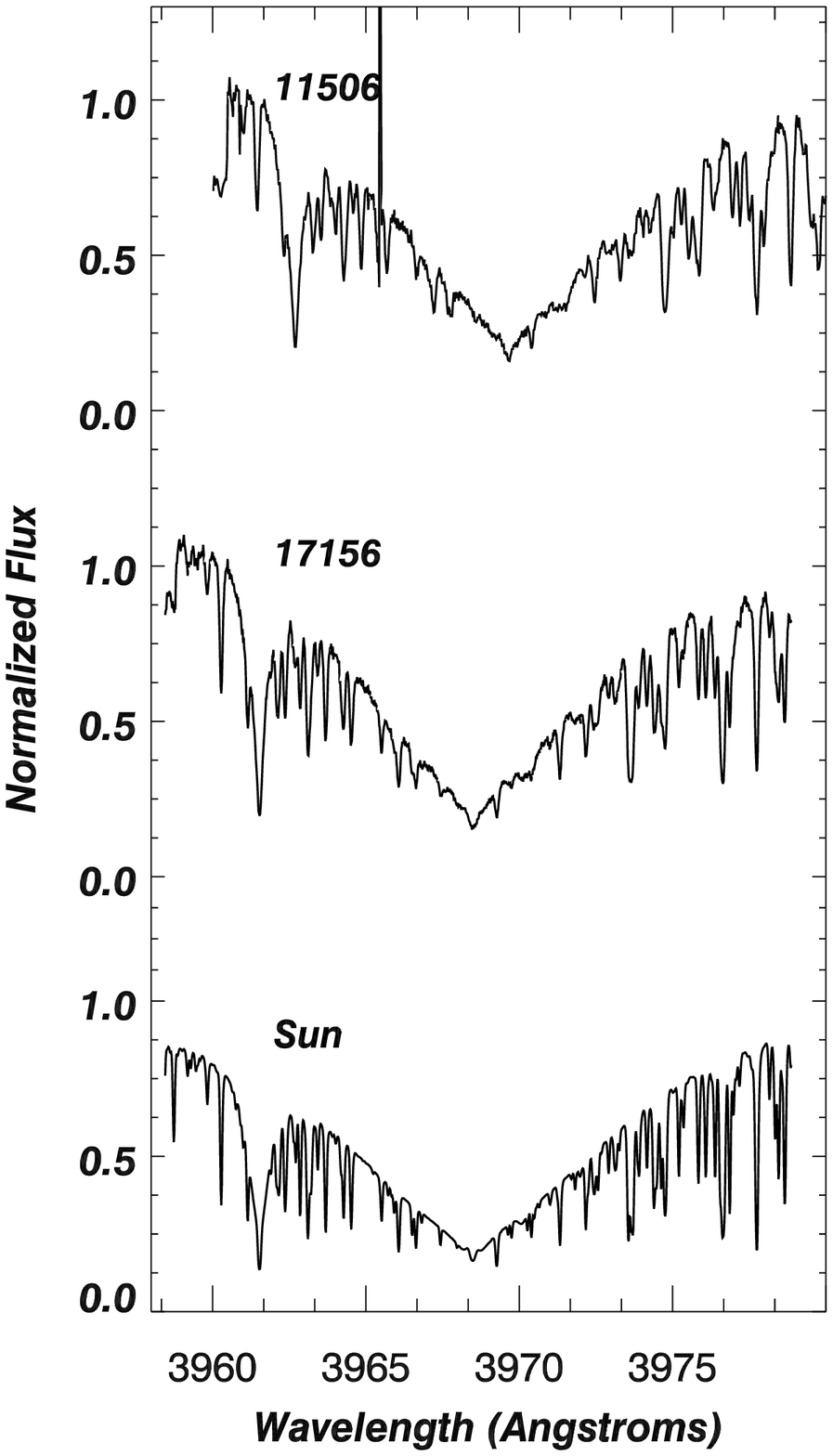}{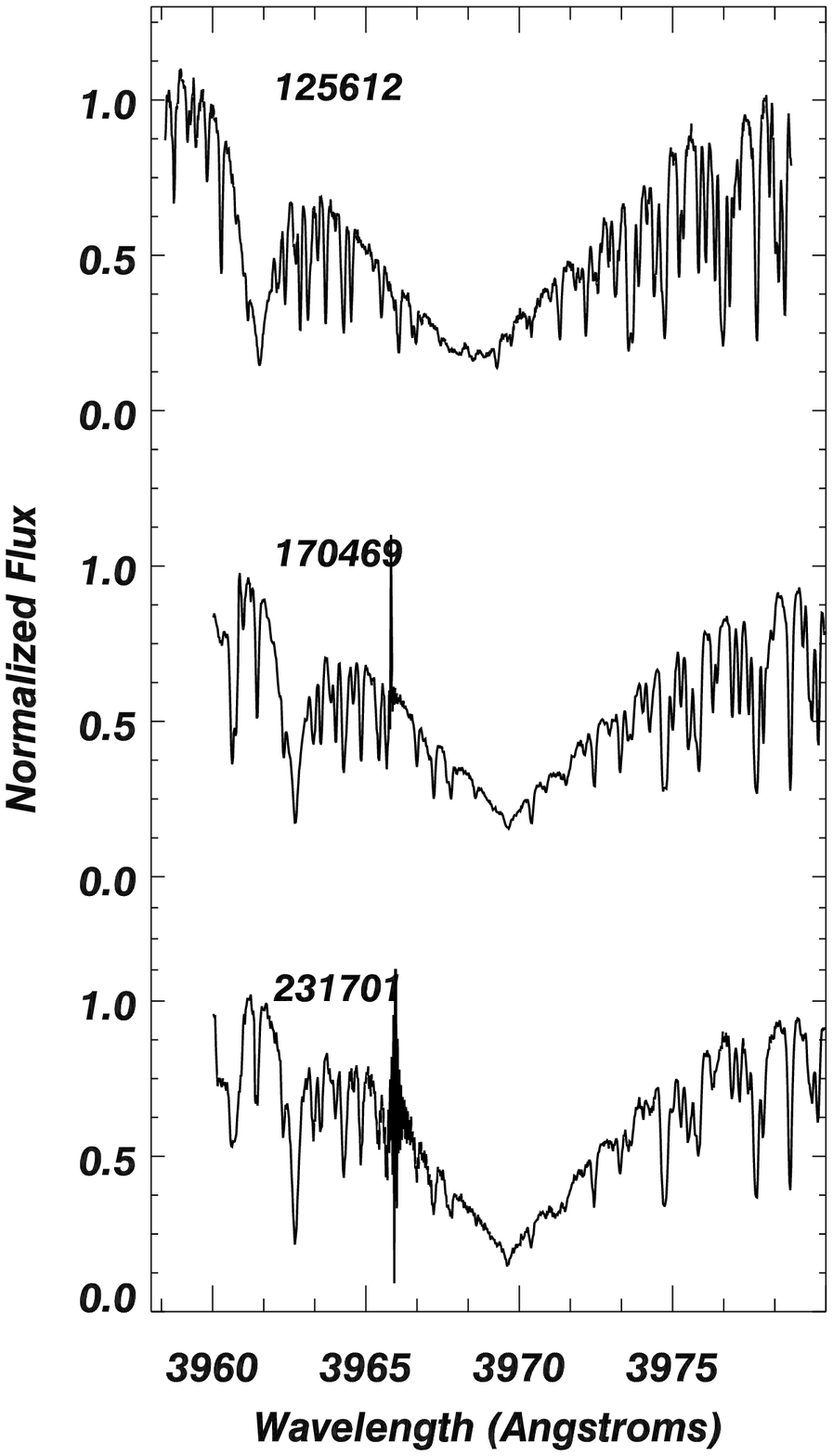}
\epsscale{1.0}
\caption{The Ca H line for HD 11506 and 17156 are plotted 
in the left panel, with the same wavelength segment of the Sun shown for 
comparison.  HD 125612, HD 170469 and HD 231701 are 
plotted in the right panel.  All of these stars have low 
chromospheric activity based on our measurement of line core emission 
relative to the continuum. }
\label{fig1}
\end{figure}
      
\begin{figure}
\plotone{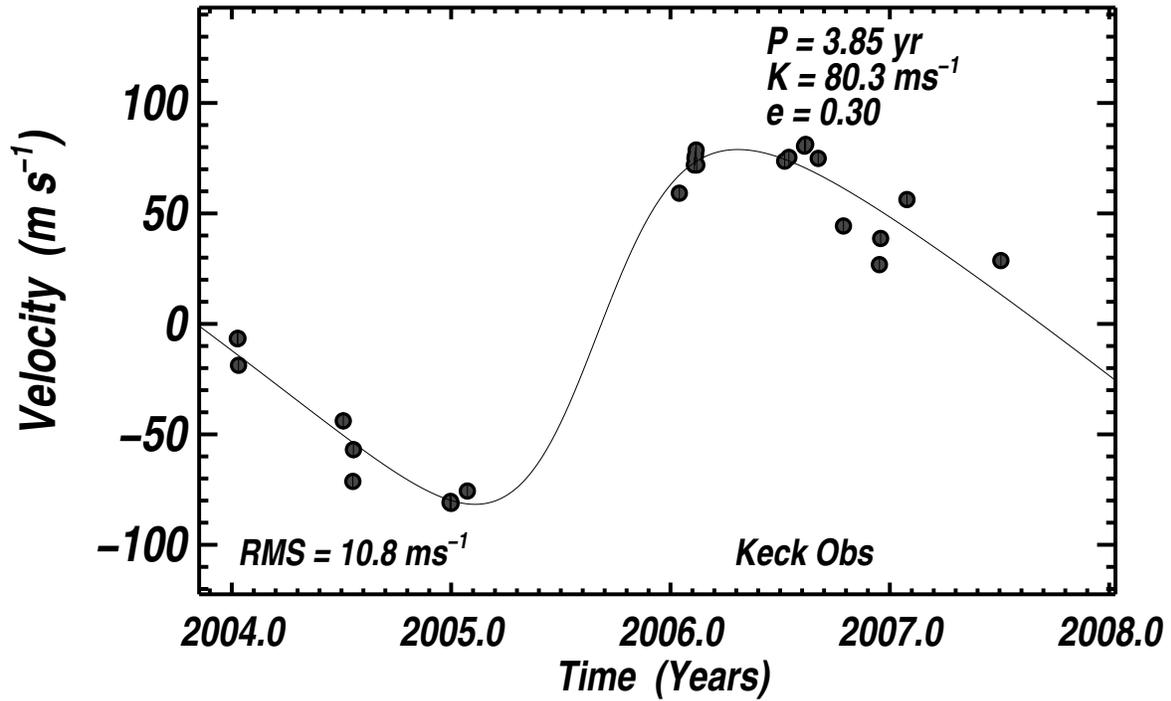}
\figcaption{Radial velocities for HD 11506.  The velocity error
bars have been augmented by adding 2 \ms in quadrature to the 
single measurement precision listed in Table 2. This gives \chisq = 3.2 
for the Keplerian fit.  With a stellar mass of 1.19 \msune, we derive 
a planet mass of \msini = 4.74 \mjup and semi-major axis, \arel = 2.48 AU. }
\label{fig2}
\end{figure}
\clearpage
  
\begin{figure}
\plottwo{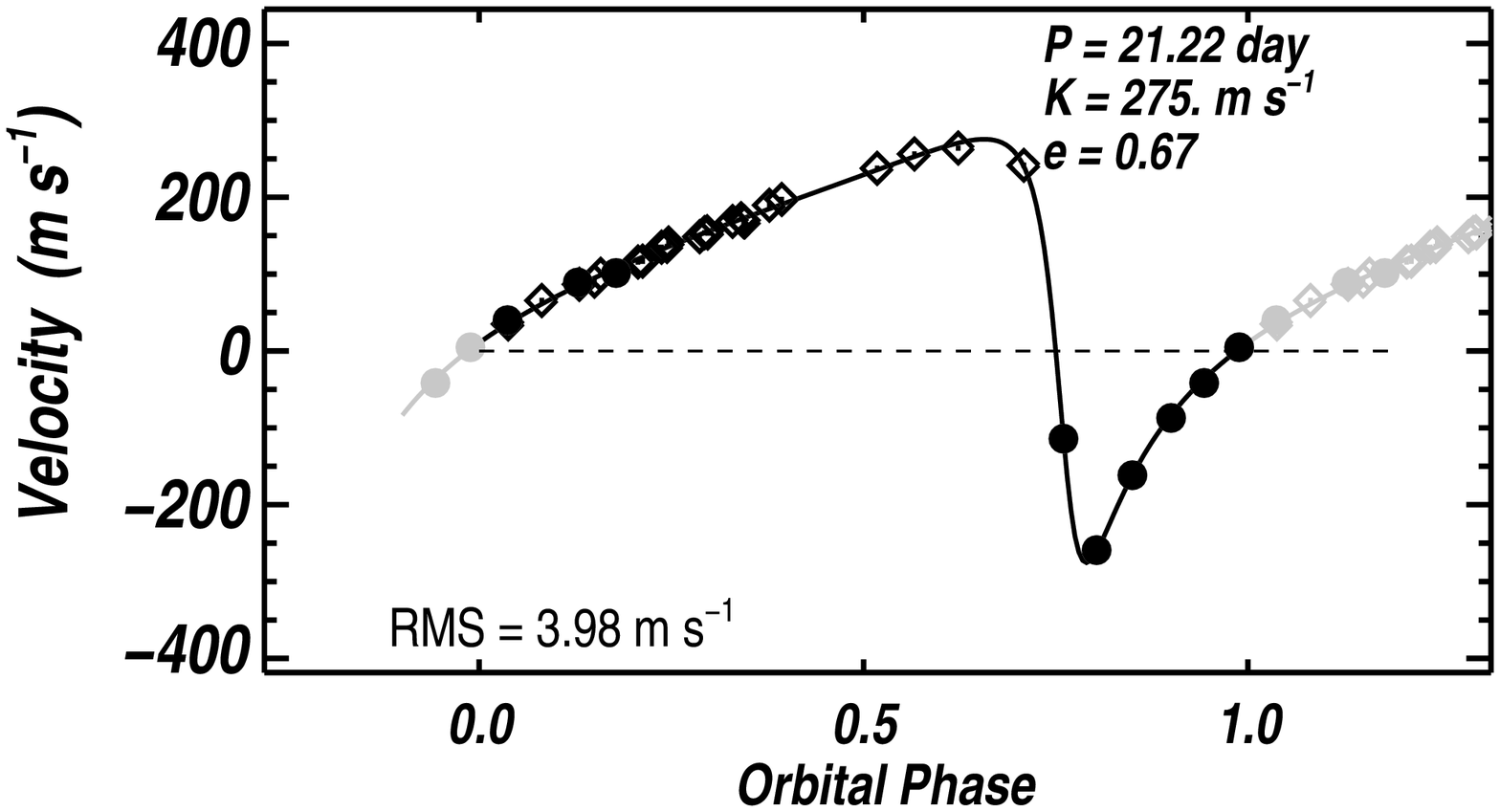}{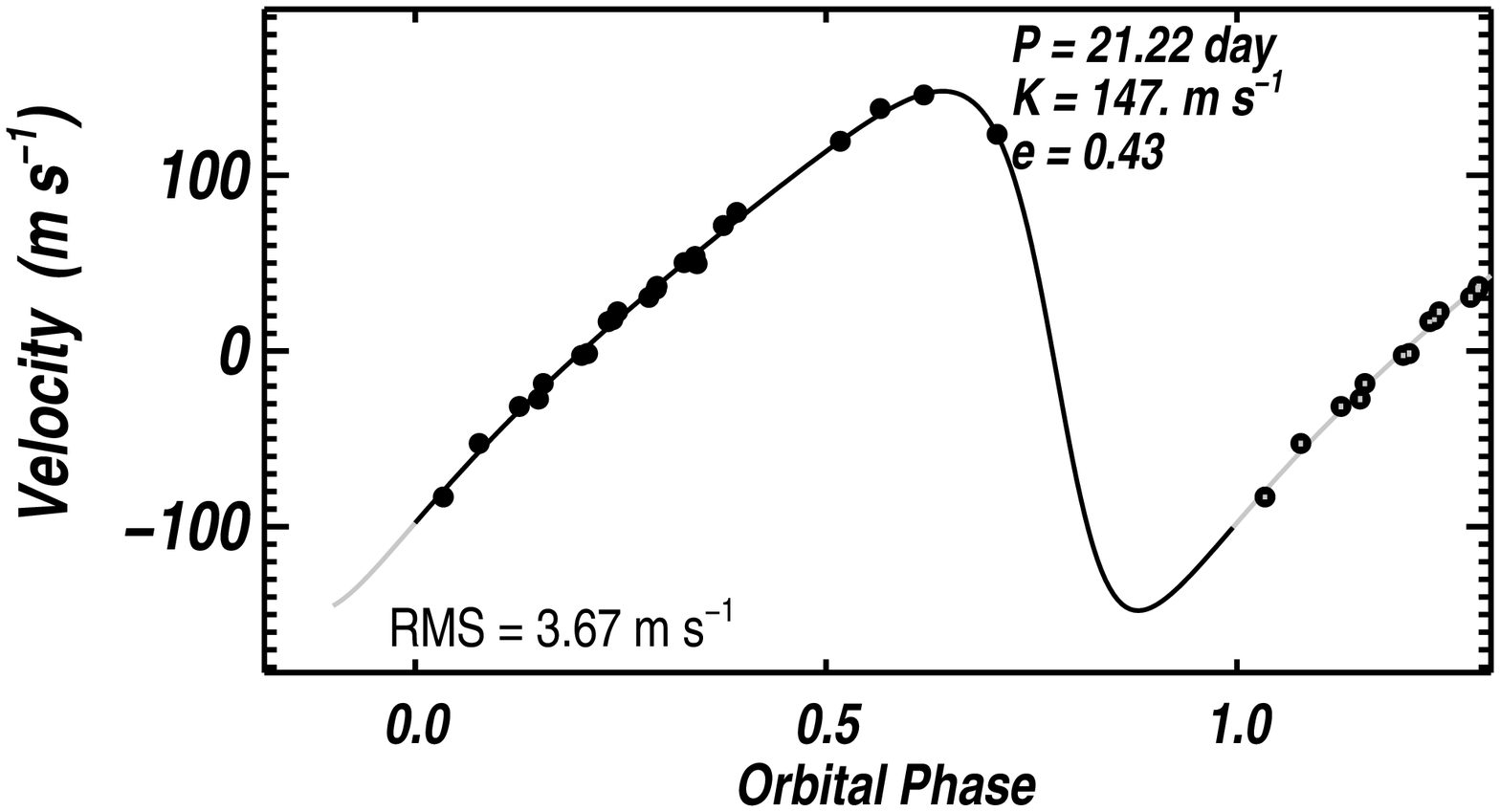}
\epsscale{1.2}
\figcaption{(left) Radial velocities for HD 17156
from Keck Observatory (diamonds) and Subaru Observatory (filled 
circles) have 3 \ms added in quadrature to 
the uncertainties listed in Table 4 
to account for expected photospheric jitter. 
Adopting a stellar mass of 1.2 \msun we derive a planet mass, \msini = 3.12 \mjup 
and semi-major axis for the orbit, \arel = 0.15 AU. 
(right) The plot on the right shows the Keck velocities only. Although 
the phase coverage misses periastron the Keck velocites alone confirm 
high eccentricity in HD 17156 b.}
\label{fig3}
\end{figure}
\clearpage
    
\begin{figure}
\plotone{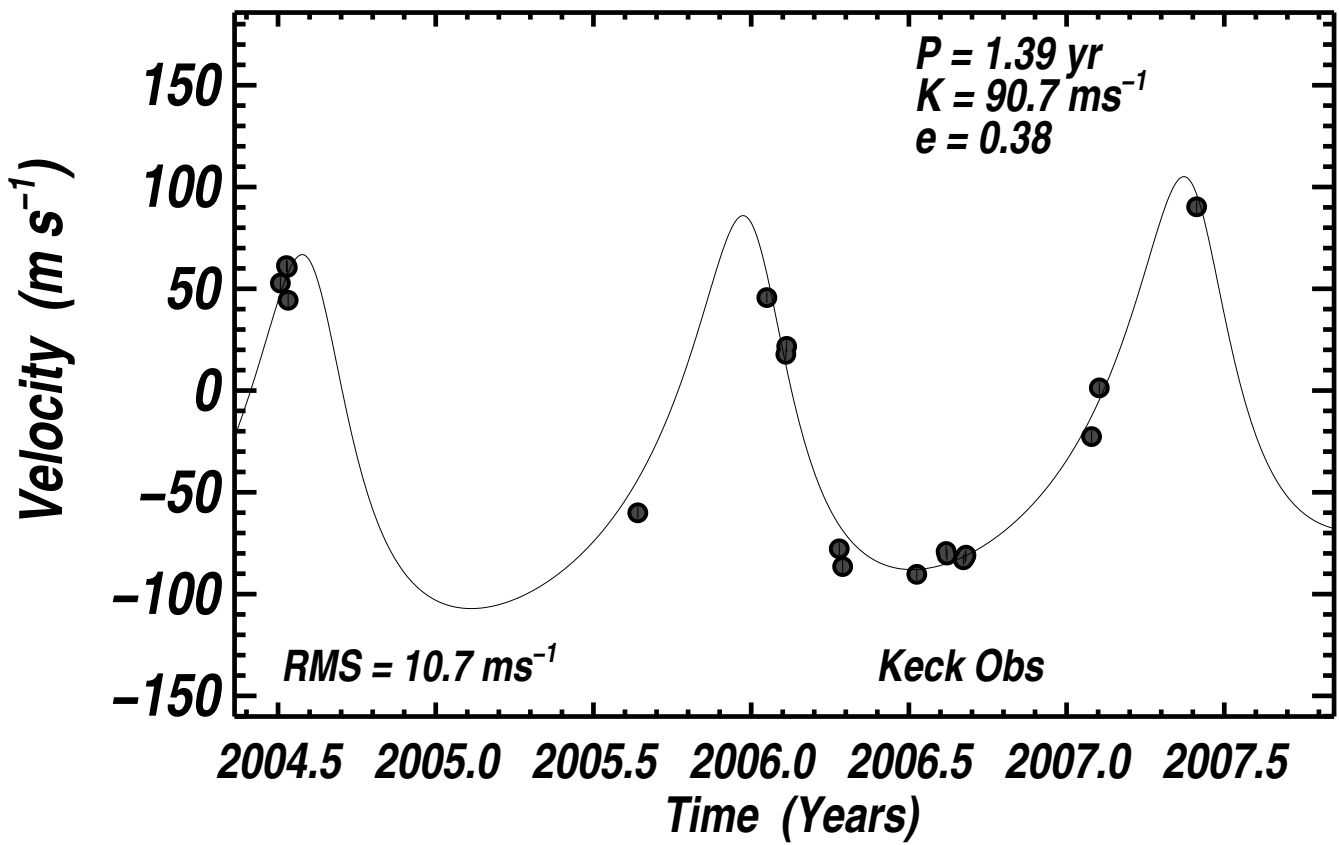}
\figcaption{Radial velocities for HD 125612.
The velocity measurements have 2 \ms added to their error bars listed 
in Table 5 to account for the level of astrophysical noise (jitter) 
we expect from the star. With the assumed stellar mass of 1.1 \msune,
we derive a planet mass, \msini = 3.5 \mjup and semi-major axis of 1.2 AU. 
This Keplerian model still has a high RMS and \chisq of 3.56, suggesting the possible 
presence of an additional planet.  }
\label{fig4}
\end{figure}
\clearpage
  
\begin{figure}
\plotone{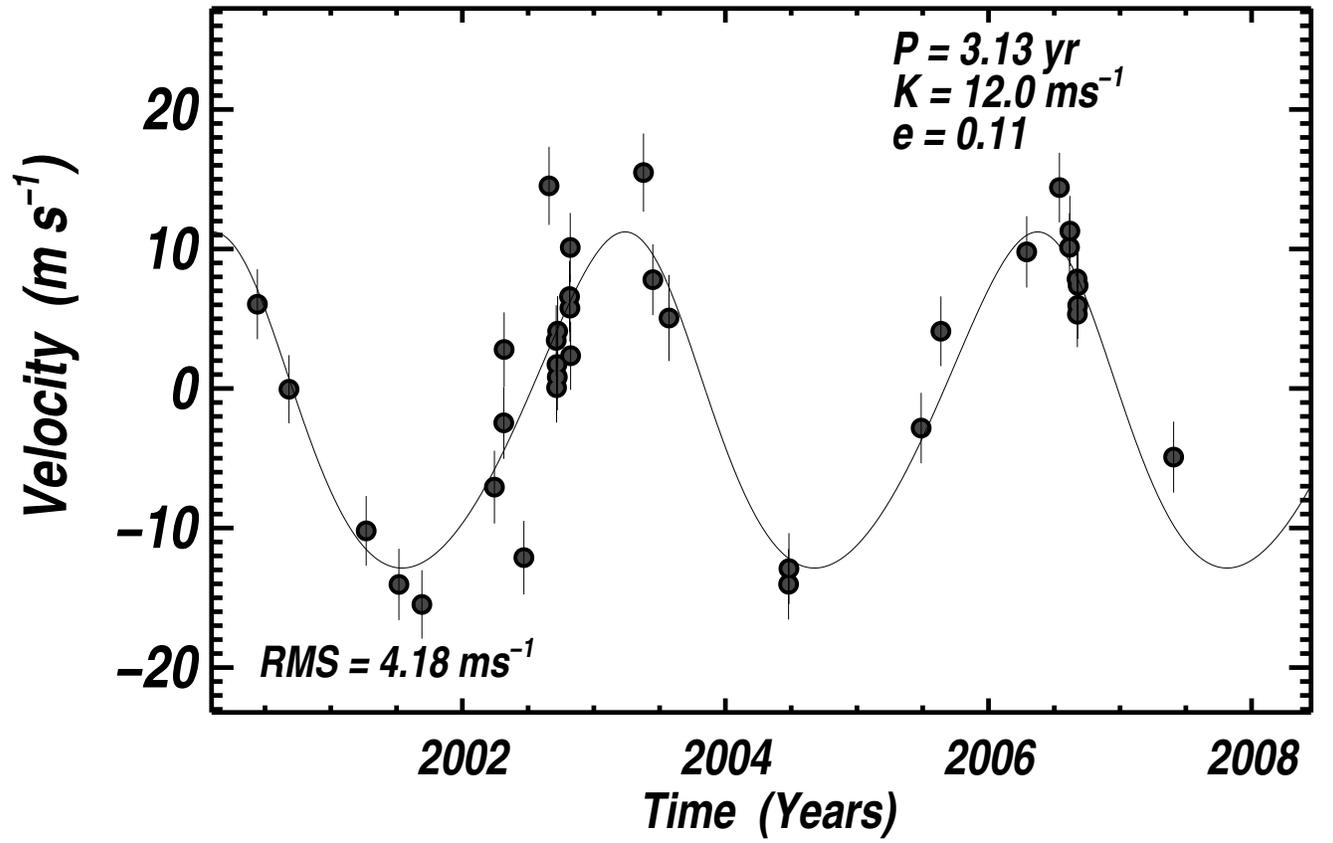}
\figcaption{Keck radial velocities for HD 170469 include 2.0 \ms 
added in quadrature with the internal error bars listed in 
Table 7. With the added jitter, the Keplerian 
fit has \chisq = 1.59.  The assumed stellar mass of 1.14 \msun yields 
a planet mass of \msini = 0.67 \mjup 
and semi-major axis of about 2 AU. }
\label{fig5}
\end{figure}
\clearpage
  
\begin{figure}
\plotone{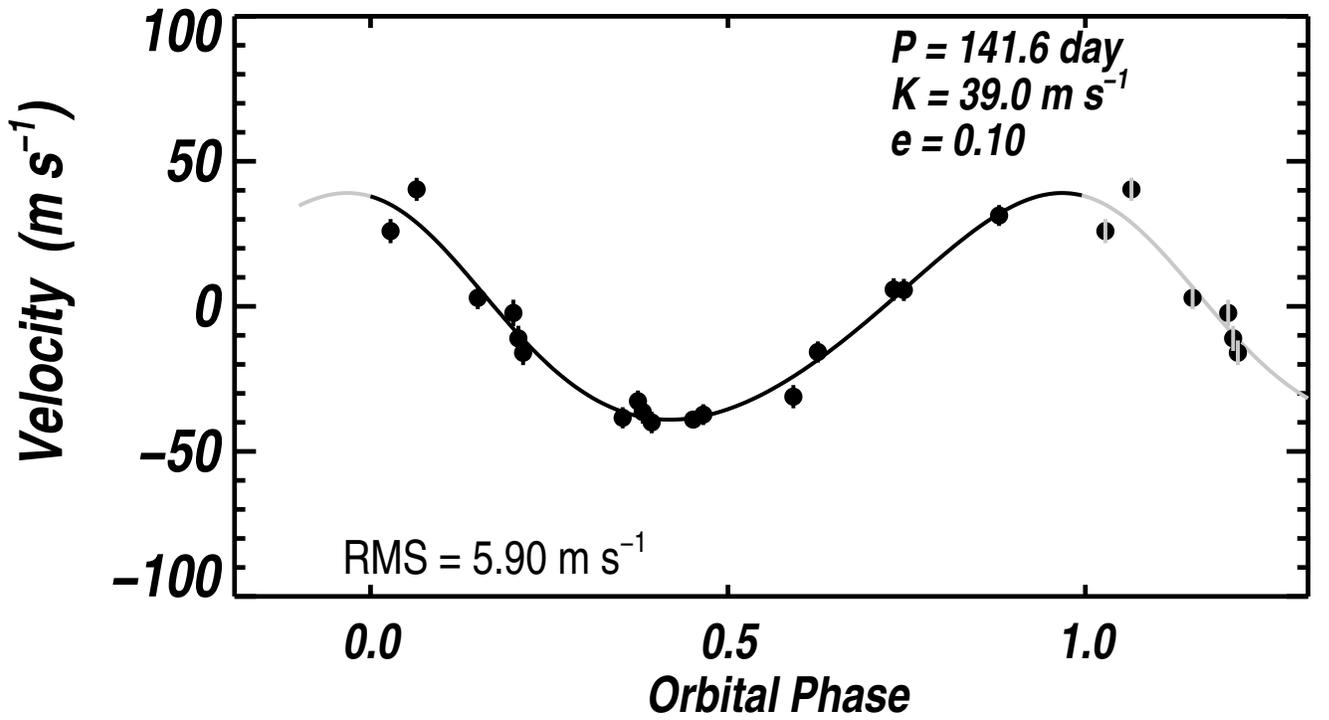}
\figcaption{Phase-folded radial velocities for HD 231701 include stellar 
jitter of 2.2 \mse added to the errors listed in Table 9, giving \chisq = 1.46.  
The stellar mass of 1.14 \msun implies a 
planet mass of \msini = 1.08 \mjup 
and orbital radius of 0.53 AU. }
\label{fig6}
\end{figure}
\clearpage

\begin{figure} 
\plotone{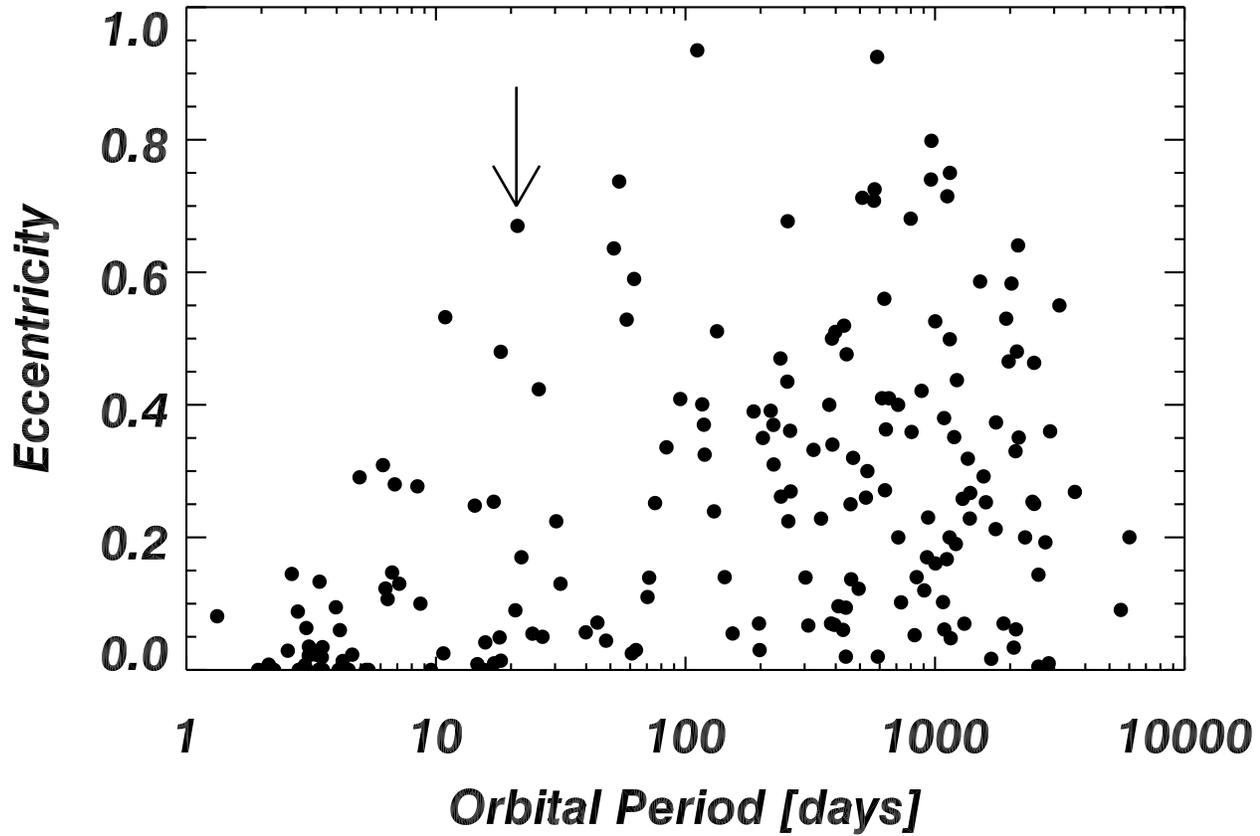}
\figcaption{Orbital eccentricity distribution for exoplanets. 
A rising envelope defines the distribution for planets with periods
between 2 and 100 days. The distribution peaks for periods between 
100 - 1000 days. The arrow points to the dot representing HD~17156 b. 
With an orbital period of 21 days and eccentricity of 
0.67, HD 17156 b still fits within the envelope of this eccentricity 
distribution.}
\label{fig7}
\end{figure}
\clearpage

\begin{figure}
\plotone{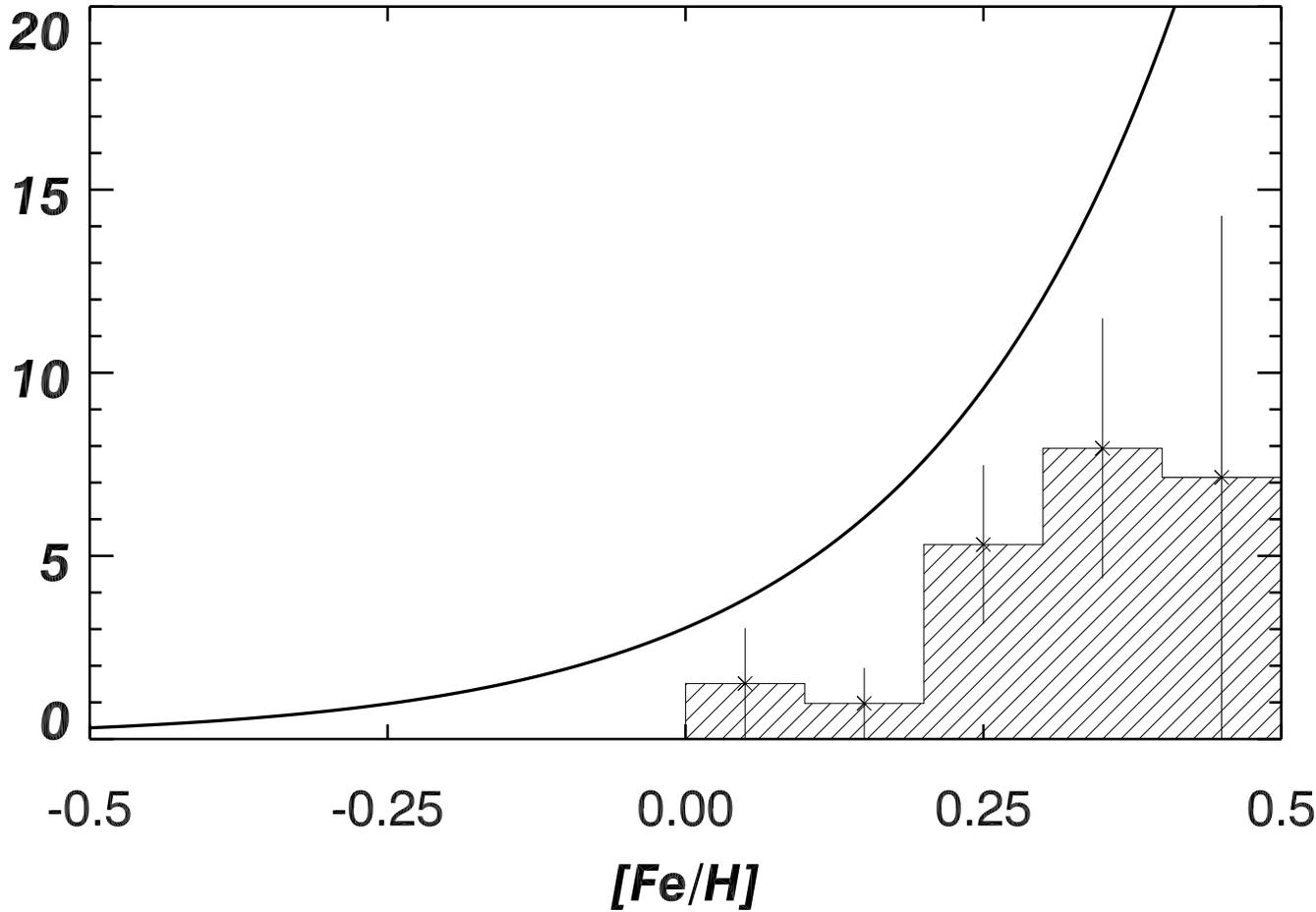}
\figcaption{Stars with at least 4 Doppler observations 
that were observed at Keck as part of the N2K program were
been binned according to metallicity. In each 0.1 dex 
metallicity bin, the percentage of stars with detected 
planets is plotted and Poisson error bars are shown. 
Superimposed on this histogram is the curve of planet 
occurrence as a function of metallicity from 
Fischer \& Valenti (2005).  Most stars on the N2K program have only  
4-5 observations, however the metallicity correlation 
is still emerging and is supported in this sample. }
\label{fig8}
\end{figure}
\clearpage

%table 1
\begin{deluxetable}{llll}
\tablenum{1}
\tablecaption{Stellar Parameters}
\tablewidth{0pt}
\tablehead{\colhead{Parameter}  & \colhead{HD11506} & \colhead{HD17156}  & \colhead{HD125612} \\
}
\startdata
V                      & 7.51               & 8.17                &    8.31            \\
$M_V$                  & 3.85               & 3.70                &    4.69            \\
B-V                    & 0.607              & 0.590               &    0.628           \\
Spectral Type          & G0V                & G0V                 &    G3V             \\
Distance (pc)          & 53.82              & 78.24               &    52.82           \\
$L_{bol}/$\lsun        & 2.29               & 2.6                 &    1.08            \\
${\rm [Fe/H]}$         & 0.31 (0.03)        & 0.24 (0.03)         &    0.24 (0.03)     \\
$T_{eff}$ (K)          & 6058 (51)          & 6079 (56)           &    5897 (40)       \\
\vsini (\kse)          & 5.0 (0.50)         & 2.6 (0.50)          &    2.1 (0.50)      \\
\logg                  & 4.32 (0.08)        & 4.29 (0.06)         &    4.45 (0.05)     \\
$M_{STAR}$ (\msun)\tablenotemark{a}  & (1.1) 1.19 (1.29)  & (1.1) 1.2 (1.3)  & (1.04) 1.1 (1.17)  \\
$R_{STAR}$ (\rsun)\tablenotemark{a}  & (1.25) 1.38 (1.53) & (1.3) 1.47 (1.6) & (0.99) 1.05 (1.13) \\
Age (Gyr)\tablenotemark{a}           & (3.9) 5.4 (7.0)    & (3.8) 5.7 (7.0)  & (0.16) 2.1 (5.6)    \\
\shk                   & 0.156              & 0.15                &     0.178           \\
\rhk                   & -4.99              & -5.04               &     -4.85           \\
\prot (d)              & 12.6 d             & 12.8 d              &     10.5 d          \\
$\sigma_{phot}$ (mag)  & 0.0023             & 0.0024              &     \nodata         \\
\enddata
\tablenotetext{a}{Stellar masses, radii and ages are derived from evolutionary tracks}
\end{deluxetable}
\clearpage

%table 2
\begin{deluxetable}{rrc}
\tablenum{2}
\tablecaption{Radial Velocities for HD~11506}
\tablewidth{0pt}
\tablehead{ \colhead{JD} & \colhead{RV} & \colhead{Uncertainties}   \\
  \colhead{-2440000.} & \colhead{(\ms)} & \colhead{(\ms)}   \\}
\startdata
    13014.73505  &     -6.67  &      2.94   \\ 
    13015.73893  &     -6.58  &      2.89   \\ 
    13016.74089  &    -18.73  &      2.98   \\ 
    13191.12201  &    -43.89  &      3.56   \\ 
    13207.10116  &    -71.30  &      3.17   \\ 
    13208.08401  &    -56.92  &      3.41   \\ 
    13368.83778  &    -80.87  &      2.26   \\ 
    13369.75897  &    -80.35  &      2.18   \\ 
    13370.73242  &    -81.24  &      2.17   \\ 
    13397.73009  &    -75.66  &      2.36   \\ 
    13750.73807  &     59.16  &      2.67   \\ 
    13775.72853  &     71.94  &      2.77   \\ 
    13776.70435  &     74.90  &      2.57   \\ 
    13777.72528  &     76.27  &      2.93   \\ 
    13778.71858  &     78.70  &      2.73   \\ 
    13779.74737  &     71.93  &      2.77   \\ 
    13926.12744  &     73.68  &      2.66   \\ 
    13933.09065  &     75.37  &      2.62   \\ 
    13959.13935  &     80.47  &      2.46   \\ 
    13961.12421  &     81.24  &      2.73   \\ 
    13981.98256  &     74.91  &      3.12   \\ 
    14023.97438  &     44.32  &      3.07   \\ 
    14083.84327  &     26.82  &      2.44   \\ 
    14085.92115  &     38.66  &      2.54   \\ 
    14129.74332  &     56.29  &      2.48   \\ 
    14286.11838  &     28.66  &      3.01   \\ 
\enddata
\end{deluxetable}
\clearpage

\clearpage

%table 3
\begin{deluxetable}{llll}
\tablenum{3}
\tablecaption{Orbital Parameters}
\tablewidth{0pt}
\tablehead{\colhead{Parameter}   & \colhead{HD 11506} & \colhead{HD 17156}  & \colhead{HD 125612}   \\
} 
\startdata
P (d)                     &  1405 (45)         &   21.2 (0.3)         &   510 (14)          \\
${\rm T}_{\rm p}$ (JD)    &  13603 (102)       &   13738.529 (0.5)    &   13228.3 (12)      \\
$\omega$ (deg)            &  262 (19)          &   121 (11)           &   21 (9)            \\
ecc                       &  0.3 (0.1)         &   0.67 (0.08)        &   0.38 (0.05)       \\
K$_1$ (\mse)              &  80 (3)            &   275 (15)           &   90.7 (8)          \\
$dv/dt$ (\mse per day)    &  \nodata           &   \nodata            &   0.037             \\
$a_{rel}$ (AU)            &  2.48              &   0.15               &   1.2               \\
$a_1 \sin i$ (AU)         &  0.0099            &   0.00039            &   0.0039            \\
f$_1$(m) (M$_\odot$)      &  6.53e-08          &   1.83e-08           &   3.09e-08          \\
$M\sin i$ (M$_{Jup}$)     &  4.74              &   3.12               &   3.5               \\
${\rm Nobs}$              &  26                &   33                 &   19                \\
RMS (\ms)                 &  10.8              &   3.97               &   10.7              \\
Jitter (\ms)              &  2                 &   3                  &   2                 \\
Reduced \chisq            &  3.2              &   1.17               &   3.56              \\
FAP (periodogram)         &  $< 0.0001$        &   $< 0.0001$         &   $0.0003$          \\
\enddata                        
\end{deluxetable}                
\clearpage

%table 4
\begin{deluxetable}{rrcc}
\tablenum{4}
\tablecaption{Radial Velocities for HD 17156}
\tablewidth{0pt}
\tablehead{ \colhead{JD} & \colhead{RV} & \colhead{Uncertainties} & \colhead{Observatory}  \\
  \colhead{-2440000.} & \colhead{(\ms)} & \colhead{(\ms)}         & \colhead{ }            \\}
\startdata
    13746.75596  &     88.49  &      1.70 & K    \\ 
    13748.79814  &    138.15  &      1.73 & K    \\ 
    13749.79476  &    151.35  &      1.67 & K    \\ 
    13750.80160  &    169.65  &      1.76 & K    \\ 
    13775.77821  &    235.17  &      1.83 & K    \\ 
    13776.80791  &    253.80  &      1.81 & K    \\ 
    13779.82897  &    239.14  &      1.64 & K    \\ 
    13959.13219  &     97.33  &      1.55 & K    \\ 
    13962.07028  &    152.64  &      1.51 & K    \\ 
    13963.10604  &    165.48  &      1.68 & K    \\ 
    13964.13118  &    194.69  &      1.70 & K    \\ 
    13982.03231  &    132.55  &      1.20 & K    \\ 
    13983.08575  &    146.35  &      1.70 & K    \\ 
    13983.99480  &    166.11  &      1.32 & K    \\ 
    13985.00847  &    187.22  &      1.57 & K    \\ 
    14023.95206  &    114.39  &      1.77 & K    \\ 
    14047.95773  &    166.01  &      1.74 & K    \\ 
    14078.01162  &   -116.60  &      5.14 & S    \\ 
    14078.92501  &   -261.56  &      5.18 & S    \\ 
    14079.91371  &   -164.10  &      5.23 & S    \\ 
    14080.98093  &    -89.57  &      5.13 & S    \\ 
    14081.89406  &    -44.08  &      5.15 & S    \\ 
    14082.86071  &      2.39  &      5.14 & S    \\ 
    14083.88445  &     37.62  &      5.16 & S    \\ 
    14083.90314  &     32.76  &      1.33 & K    \\ 
    14084.82860  &     63.22  &      1.63 & K    \\ 
    14085.82560  &     86.67  &      5.20 & S    \\ 
    14085.86537  &     84.28  &      1.56 & K    \\ 
    14086.87960  &     99.01  &      5.20 & S    \\ 
    14129.92513  &    113.30  &      1.43 & K    \\ 
    14130.73019  &    133.66  &      1.32 & K    \\ 
    14131.85485  &    151.11  &      1.77 & K    \\ 
    14138.76720  &    261.56  &      1.37 & K    \\ 
\enddata
\end{deluxetable}
\clearpage

\clearpage
 
%table 5
\begin{deluxetable}{rrc}
\tablenum{5}
\tablecaption{Radial Velocities for HD 125612}
\tablewidth{0pt}
\tablehead{ \colhead{JD} & \colhead{RV} & \colhead{Uncertainties}   \\
  \colhead{-2440000.} & \colhead{(\ms)} & \colhead{(\ms)}   \\}
\startdata
    13190.83262  &     52.80  &      2.85   \\ 
    13197.83363  &     61.40  &      2.66   \\ 
    13198.85557  &     60.49  &      2.59   \\ 
    13199.83792  &     44.35  &      2.70   \\ 
    13604.75480  &    -60.07  &      2.04   \\ 
    13754.12824  &     45.68  &      1.83   \\ 
    13776.15380  &     17.80  &      2.03   \\ 
    13777.11940  &     21.71  &      2.16   \\ 
    13838.01492  &    -77.76  &      2.35   \\ 
    13841.94894  &    -86.49  &      2.52   \\ 
    13927.79007  &    -90.28  &      2.01   \\ 
    13961.75080  &    -79.07  &      2.01   \\ 
    13962.74127  &    -80.81  &      1.83   \\ 
    13981.72862  &    -83.16  &      1.90   \\ 
    13983.74037  &    -81.92  &      2.11   \\ 
    13984.72815  &    -80.88  &      2.03   \\ 
    14130.13392  &    -22.62  &      2.18   \\ 
    14139.12383  &      1.31  &      1.94   \\ 
    14251.82778  &     90.28  &      2.39   \\ 
\enddata
\end{deluxetable}
\clearpage

\clearpage

\begin{deluxetable}{lll}
\tablenum{6}
\tablecaption{Stellar Parameters}
\tablewidth{0pt}
\tablehead{\colhead{Parameter}  &  \colhead{HD 170469} &  \colhead{HD 231701}   \\
}
\startdata
V                      & 8.21               & 8.97                   \\
$M_V$                  & 4.14               & 3.79                   \\
B-V                    & 0.677              & 0.539                  \\
Spectral Type          & G5IV               & F8V                    \\
Distance (pc)          & 64.97              & 108.4                  \\
$L_{bol}/$\lsun        & 1.6                & 2.4                    \\
${\rm [Fe/H]}$         & 0.30 (0.03)        & 0.07 (0.03)            \\
$T_{eff}$ (K)          & 5810 (44)          & 6208 (44)              \\
\vsini (\kse)          & 1.7 (0.5)          & 4 (0.50)               \\
\logg                  & 4.32 (0.06)        & 4.33 (0.06)            \\
$M_{STAR}$ (\msun)\tablenotemark{a}     & (1.05) 1.14 (1.16) & (1.08) 1.14 (1.22)     \\
$R_{STAR}$ (\rsun)\tablenotemark{a}     & (1.15) 1.22 (1.3)  & (1.16) 1.35 (1.55)     \\
Age (Gyr)\tablenotemark{a}              & (5.0) 6.7 (7.8)    & (3.5) 4.9 (6.2)        \\
\shk                   & 0.145              & 0.159                  \\
\rhk                   & -5.06              & -5.00                  \\
\prot (d)              & 13.0 d             & 12.2 d                 \\
$\sigma_{phot}$ (mag)  & 0.0018             & \nodata                \\
\enddata
\tablenotetext{a}{Stellar masses, radii and ages are derived from evolutionary tracks}
\end{deluxetable}
\clearpage

%table 7
\begin{deluxetable}{rrc}
\tablenum{7}
\tablecaption{Radial Velocities for HD~170469}
\tablewidth{0pt}
\tablehead{ \colhead{JD} & \colhead{RV} & \colhead{Uncertainties}   \\
  \colhead{-2440000.} & \colhead{(\ms)} & \colhead{(\ms)}   \\}
\startdata
    11705.96808  &      6.04  &      1.51   \\ 
    11793.81330  &     -0.05  &      1.39   \\ 
    12008.04881  &    -10.21  &      1.50   \\ 
    12099.03294  &    -14.05  &      1.59   \\ 
    12162.76894  &    -15.48  &      1.42   \\ 
    12364.13287  &     -7.07  &      1.67   \\ 
    12390.12499  &     -2.46  &      1.63   \\ 
    12391.12567  &      2.80  &      1.76   \\ 
    12445.93867  &    -12.12  &      1.72   \\ 
    12515.82777  &     14.53  &      1.96   \\ 
    12535.75539  &      3.45  &      1.53   \\ 
    12536.74191  &      0.06  &      1.50   \\ 
    12537.82520  &      1.72  &      1.50   \\ 
    12538.74254  &      0.83  &      1.30   \\ 
    12539.75501  &      4.11  &      1.51   \\ 
    12572.69435  &      6.59  &      1.60   \\ 
    12573.69333  &      5.76  &      1.34   \\ 
    12574.70725  &     10.11  &      1.45   \\ 
    12575.69822  &      2.35  &      1.40   \\ 
    12778.04455  &     15.48  &      1.96   \\ 
    12804.05044  &      7.80  &      1.56   \\ 
    12848.92274  &      5.06  &      2.35   \\ 
    13180.90825  &    -14.03  &      1.55   \\ 
    13181.89752  &    -12.91  &      1.57   \\ 
    13548.99248  &     -2.83  &      1.54   \\ 
    13603.80234  &      4.11  &      1.50   \\ 
    13842.01212  &      9.79  &      1.59   \\ 
    13932.96654  &     14.41  &      1.49   \\ 
    13960.91798  &     10.12  &      1.42   \\ 
    13961.83115  &     11.29  &      1.54   \\ 
    13981.82421  &      7.85  &      1.27   \\ 
    13982.77494  &      5.33  &      1.26   \\ 
    13983.76067  &      5.96  &      1.33   \\ 
    13984.83377  &      7.38  &      1.22   \\ 
    14250.01196  &     -4.92  &      1.57   \\ 
\enddata
\end{deluxetable}
\clearpage

\clearpage
 
\begin{deluxetable}{lll}
\tablenum{8}
\tablecaption{Orbital Parameters}
\tablewidth{0pt}
\tablehead{\colhead{Parameter}   & \colhead{HD 170469} & \colhead{HD 231701}  \\
} 
\startdata
P (d)                     &    1145 (18)          &   141.6 (2.8)          \\
${\rm T}_{\rm p}$ (JD)    &    11669.0 (21)       &   13180.0 (4.2)        \\
$\omega$ (deg)            &    34 (19)            &   46 (24)               \\
ecc                       &    0.11 (0.08)        &   0.10 (0.08)          \\
K$_1$ (\ms)               &    12.0 (1.9)         &   64 (8)               \\
$dv/dt$ (\mse per day)    &    \nodata            &   \nodata              \\
$a_{rel}$ (AU)            &    2.1                &   0.53                 \\
$a_1 \sin i$ (AU)         &    0.00126            &   0.0005               \\
f$_1$(m) (M$_\odot$)      &    2.03e-10           &   8.6e-10               \\
$M\sin i$ (M$_{Jup}$)     &    0.67               &   1.08                  \\
${\rm Nobs}$              &    35                 &   17                   \\
RMS (\ms)                 &    4.18               &   5.90                 \\
Jitter (\ms)              &    2.0                &   2.22                 \\
Reduced \chisq            &    1.59               &   1.46                 \\
FAP (periodogram)         &    $< 0.0001$         &   0.006                \\
\enddata                        
\end{deluxetable}                
\clearpage

%table 9
\begin{deluxetable}{rrc}
\tablenum{9}
\tablecaption{Radial Velocities for HD~231701}
\tablewidth{0pt}
\tablehead{ \colhead{JD} & \colhead{RV} & \colhead{Uncertainties}   \\
  \colhead{-2440000.} & \colhead{(\ms)} & \colhead{(\ms)}   \\}
\startdata
    13190.98047  &      1.98  &      3.41   \\ 
    13198.03808  &     -2.44  &      4.00   \\ 
    13199.02309  &    -11.19  &      3.74   \\ 
    13199.95689  &    -16.16  &      3.62   \\ 
    13603.87134  &     40.18  &      3.32   \\ 
    13928.04498  &    -38.57  &      2.90   \\ 
    13931.08095  &    -32.84  &      2.93   \\ 
    13932.02123  &    -36.55  &      3.38   \\ 
    13961.87814  &    -31.29  &      3.34   \\ 
    13981.76292  &      5.66  &      3.18   \\ 
    13983.77988  &      5.49  &      3.11   \\ 
    14023.71873  &     25.79  &      3.53   \\ 
    14083.69519  &    -39.18  &      2.43   \\ 
    14085.70009  &    -37.48  &      2.83   \\ 
    14217.13510  &    -40.18  &      2.99   \\ 
    14250.07410  &    -15.89  &      2.90   \\ 
    14286.00169  &     31.17  &      2.86   \\ 
\enddata
\end{deluxetable}
\clearpage

\clearpage

\end{document}